\documentclass[pre,twocolumn,showpacs]{revtex4}
\usepackage{epsfig}

\begin{document}

\title{Numerical modeling of quasiplanar giant water waves}
\author{Victor P. Ruban$^{1}$}
\email{ruban@itp.ac.ru}
\author{J\"urgen Dreher$^{2}$}
\email{dreher@tp1.rub.de}
\affiliation{$^{1}$Landau Institute for Theoretical Physics,
2 Kosygin Street, 119334 Moscow, Russia} 
\affiliation{$^{2}$Theoretische Physik I, Ruhr-Universit\"at Bochum, Germany}

\date{\today}

\begin{abstract}
In this work we present a further analytical development and a numerical
implementation of the recently suggested theoretical model for highly 
nonlinear potential long-crested water waves, where weak three-dimensional 
effects are included as small corrections to exact two-dimensional equations 
written in the conformal variables 
[V.P. Ruban, Phys. Rev. E {\bf 71}, 055303(R) (2005)].
Numerical experiments based on this theory describe the spontaneous 
formation of a single weakly three-dimensional large-amplitude wave 
(alternatively called freak, killer, rogue or giant wave) on the deep water.
\end{abstract}

\pacs{47.15.Hg, 02.60.Cb, 92.10.-c}
\maketitle


\section{Introduction}

Rogue waves are extremely high, steep and dangerous individual waves which 
sometimes appear suddenly on a sea surface among relatively low waves (see, 
for instance, the recent works \cite{Kharif-Pelinovsky,ZDV2002,DZ2005Pisma}, 
and references therein). The crest of a rogue wave can be three 
or even more times higher than the crests of neighboring waves.
Different physical mechanisms contribute to the rogue wave phenomenon:
dispersion enhancement, geometrical focusing, wave-current interaction 
\cite{Kharif-Pelinovsky}, 
but the most important at the final stage is the nonlinear self-focusing
mechanism resulting in accumulation of the wave energy and momentum on 
the scale of a single wavelength \cite{DZ2005Pisma}.
For weakly modulated periodic planar Stokes waves this mechanism leads to 
the well-known Benjamin-Feir instability \cite{Benjamin-Feir,Zakharov67}, 
generated by four-wave nonlinear resonant interactions $2\to 2$.
This instability is predominantly two-dimensional (2D), and it is dominant 
for low amplitudes ($h/\lambda\le 0.09$ where  $h$ is the 
peak-to-trough height and $\lambda$ is the length of the Stokes wave 
\cite{McLean_et_al_1981}). For larger steepness parameter $h/\lambda$, 
another, genuinely three-dimensional (3D), instability becomes dominant, 
which is generated by five-wave interactions $3\to 2$, and results in the 
well-known crescent, or ``horse-shoe'' wave patterns 
(see \cite{SBK1996,Collard-Caulliez-1999,Dias-Kharif}, and references therein). 
It is important that real ocean giant waves are observed in situations 
when this 3D instability is not principal, and all the waves are typically 
long-crested, corresponding to a narrow-angle Fourier spectrum. 
Thus, many essential features of a rogue wave formation
can be observed already in purely 2D geometry, as in the works by
Zakharov and co-workers \cite{ZDV2002,DZ2005Pisma}. For instance, 
in Ref.~\cite{DZ2005Pisma} a numerical giant wave was computed with 
the impressive spatial resolution of up to $2\cdot 10^6$ points.
Zakharov and co-workers simulated an 
exact system of dynamic equations for 2D free-surface inviscid potential flows, 
written in terms of the so called conformal variables, which make the free
boundary effectively flat (the corresponding exact 2D theory is described in
Refs.~\cite{DKSZ96,DZK96,ZD96,DLZ95,Lvov97,D2001,CC99}).
With these variables, highly nonlinear equations of motion for planar 
water waves are represented in an exact and compact form containing 
integral operators diagonal in the Fourier representation. 
Such integro-differential equations are easy to treat numerically
with modern libraries for the discrete fast Fourier transform (FFT)
as, for example,  FFTW \cite{fftw3}. 
Recently, by introducing an additional conformal mapping,
the exact 2D conformal description has been generalized to 
non-uniform and time-dependent bottom profiles, so that a very accurate 
2D modeling of near-coastal waves and tsunami-like processes has 
been possible \cite{R2004PRE,R2005PLA}.

However, real sea waves are never ideally planar, and the second horizontal 
dimension might play an important role in the wave dynamics. Various numerical 
methods have been developed for nonlinear 3D surface gravity waves (see 
\cite{TY1996ARFM,Dias-Kharif,Dias-Bridges} for a review). 
Some of them are based on exact formulation of the problem (the boundary 
integral method and its modifications; see 
\cite{L-H_C-1976,Clamond-Grue-2001,Fructus_et_al_2005} and references 
therein), another approach uses approximate equations of motion, as
the Boussinesq-type models \cite{Kirby,BinghamAgnon}, 
the equations derived by Matsuno \cite{Matsuno}, by Choi \cite{Choi95},
the weakly nonlinear Zakharov equations \cite{OOSRPZB2002,DKZ2004},
or the equations for wave packets --- the nonlinear Schroedinger 
equation (NLS) and its extensions
\cite{Dysthe1979,TKDV2000,OOS2000,Janssen2003}. 
The numerical methods based on exact equations
are quite ``expensive'' and thus provide a relatively low spatial resolution 
(typically $128\times 64$, as in the recent work \cite{Fructus_et_al_2005}
for essentially 3D waves).
On the other hand, the applicability of the approximate equations
is limited by the condition that the waves must not be too steep.
To fill this gap, some new approximate, relatively compact explicit equations 
of motion for highly nonlinear 3D waves were needed as the basis for a new 
numerical method. Recently, as an extension of the exact conformal 
2D theory, a weakly 3D conformal theory has been suggested \cite{R2005PRE}, 
which is valid for steep long-crested waves. Equations of this theory contain 
3D corrections of the order of $\epsilon=(l_x/l_q)^2$, where $l_x$ is a typical 
wave length, and $l_q$ is a large transversal scale along the wave crests.
In Ref.~\cite{R2005PRE}, the general case was considered, with a static 
nonuniform quasi-one-dimensional bottom profile. Since the corresponding 
general equations are rather involved, it is natural to focus on 
some simple particular cases in the first place, for instance on the deep 
water limit. 
The purpose of the present work is a further development of the weakly 3D
conformal theory and its numerical implementation for the most simple deep 
water case. Our main result here is that we have developed a
numerical scheme based on FFT which is sufficiently simple, 
accurate and fast simultaneously. 
As an application, we have simulated a freak wave formation with 
the final resolution up to $16384\times 256$.

This article is organized as follows. In Sec.~II we adopt for the deep water 
limit the general weaky 3D fully nonlinear theory described in 
Ref.~\cite{R2005PRE}. We also suggest some modification of the obtained
dynamical system which results in the  correct linear dispersion relation 
not only within a small region near the $k_1$-axis, but in the whole 
Fourier-plane. In Sec.~III we describe our numerical method and present
numerical results for the problem of a rogue wave formation. 
Finally, Sec.~IV contains our summary and discussion. 

\section{Weakly 3D nonlinear theory}

\subsection{General remarks}

It is a well known fact that a principal difficulty in the 3D theory
of potential water waves is the general impossibility to solve exactly 
the Laplace equation for the velocity potential $\varphi(x,y,q,t)$,
\begin{equation}\label{Laplace_xyq}
\varphi_{xx}+\varphi_{yy}+\varphi_{qq}=0,
\end{equation}
in the flow region $-\infty \le y\le \eta(x,q,t)$, 
with the given boundary conditions
\begin{equation}\label{boundary_cond_Laplace_xyq}
\varphi|_{y=\eta(x,q,t)}= \psi(x,q,t), \quad \varphi|_{y=-\infty}=0.
\end{equation}
(Here $x$ and $q$ are the horizontal Cartesian coordinates, $y$ is the vertical
coordinate, while the symbol $z$ will be used for the complex combination
$z=x+iy$). Therefore a compact expression is absent for the Hamiltonian
functional of the system,
\begin{eqnarray}
{\cal H}\{\eta,\psi\}&=&\frac{1}{2}\int dx\, dq
\int\limits_{-\infty}^{\eta(x,q,t)}(\varphi_x^2+\varphi_y^2+\varphi_q^2)dy
\nonumber\\
&&+\frac{g}{2}\int\eta^2dx\,dq\equiv {\cal K}\{\eta,\psi\}
+{\cal P}\{\eta\},
\label{Hamiltonian_eta_psi}
\end{eqnarray} 
(the sum of the kinetic energy of the fluid and the potential energy in
the vertical gravitational field $g$). The Hamiltonian determines canonical 
equations of motion (see \cite{Z1999,RR2003PRE,DKZ2004}, 
and references therein)
\begin{equation}\label{Hamiltonian_equations_eta_psi}
 \eta_t=({\delta{\cal H}}/{\delta\psi}),\qquad
-\psi_t=({\delta{\cal H}}/{\delta\eta})
\end{equation}
in accordance with the variational principle $\delta\int \tilde{\cal L}dt=0$,
where the Lagrangian is
$\tilde{\cal L}=\int\psi\eta_t \,dx\,dq-{\cal H}.$

In the traditional approach, the problem is partly solved by an asymptotic 
expansion of the kinetic energy ${\cal K}$ on a small parameter --- the
steepness of the surface (see Refs. \cite{Z1999,DKZ2004,LZ2004}, 
and references therein). 
As the result, a weakly nonlinear theory is generated, which is not good to 
describe large-amplitude steep waves [see Ref. \cite{LZ2004} for a discussion 
about the limits of such theory; practically, the wave steepness $ka$
($k$ is a wave number, $a$ is an amplitude)
should be not more than $0.1$ (that is $h/\lambda\le 0.1/\pi\approx 0.03$)].
Below in this section we consider a different, recently developed theory 
\cite{R2005PRE} (adopted here for the case of infinite depth), 
which is based on another small parameter --- 
the ratio of a typical length of the waves propagating along the
$x$-axis, to a large scale along the transversal horizontal direction,
denoted by $q$ [alternatively, it is the ratio of typical wave numbers 
$k_q/k_x$ in the Fourier plane $(k_x,k_q)$]. 
Thus, we define $\epsilon=(l_x/l_q)^2\ll 1$ and note: 
the less this parameter, the less our flow differs from a purely 2D flow,
and the more accurate the equations are.
A profile $y=\eta(x,q,t)$ of the free surface and a boundary value of the
velocity potential $\psi(x,q,t)\equiv\varphi(x,\eta(x,q,t),q,t)$
are allowed to depend strongly on the coordinate $x$,
while the derivatives over the coordinate $q$ will be small:
$|\eta_q|\sim\epsilon^{1/2}$, $|\psi_q|\sim\epsilon^{1/2}$.

\subsection{Conformal variables in 3D}

In the same manner as in the exact 2D theory \cite{DKSZ96,DLZ95},
instead of the Cartesian coordinates $x$ and $y$, we use curvilinear conformal
coordinates  $u$ and $v$, which make the free surface effectively flat:
\begin{equation}\label{conformal_mapping}
x+iy\equiv z=z(u+iv,q,t), \quad -\infty\le v\le 0,
\end{equation}
where $z(w,q,t)$ is an analytical on the complex variable $w\equiv u+iv$
function without any singularities in the lower half-plane $-\infty\le v\le 0$.
The profile of the free surface is now given in a parametric form 
by the formula
\begin{equation}\label{surface_profile}
X(u,q,t)+iY(u,q,t)\equiv Z(u,q,t)=z(u+0i,q,t).
\end{equation}
Here the real functions  $X(u,q,t)$ and  $Y(u,q,t)$ are related to each other
by the Hilbert transform $\hat H$: 
\begin{equation}\label{X}
X(u,q,t)=u-\hat H Y(u,q,t).
\end{equation}
The Hilbert operator $\hat H$ is diagonal in the Fourier representation:
it multiplies the Fourier-harmonics 
$Y_k(q,t)\equiv\int Y(u,q,t)e^{-iku}du$ by $i\,\mbox{sign\,}k$,  
so that
\begin{equation}\label{HY}
\hat H Y(u,q,t)=\int [i\,\mbox{sign\,}k] Y_k(q,t)e^{iku}({dk}/{2\pi}).
\end{equation}
The boundary value of the velocity potential is 
$\varphi|_{v=0}\equiv \psi(u,q,t)$. 

For equations to be shorter, below we do not indicate the arguments $(u,q,t)$
of the functions $\psi$, $Z$ and $\bar Z$ (the overline denotes complex
conjugate). 
The Lagrangian for 3D deep water waves in terms of the variables $\psi$, $Z$, 
and $\bar Z$ can be written as follows (compare with \cite{R2005PRE}):
\begin{eqnarray}
{\cal L}&=&\int \left[
\frac{Z_t\bar Z_u -\bar Z_t Z_u}{2i}\right]\psi \,du\,dq 
-{\cal K}\{\psi,Z,\bar Z\}\nonumber\\
&-&\frac{g}{2}\int\left[\frac{Z-\bar Z}{2i}\right]^2
\left[\frac{Z_u+\bar Z_u}{2}\right]du\,dq
\nonumber\\
\label{Lagrangian_Z_psi}
&+&\int\Lambda\left[\hat H\left(\frac{Z-\bar Z}{2i}\right)
+\left(\frac{Z+\bar Z}{2}-u\right)\right]du\,dq,
\end{eqnarray}
where the indefinite real Lagrangian multiplier $\Lambda(u,q,t)$ has been
introduced in order to take into account the relation (\ref{X}).
Equations of motion follow from the variational principle 
$\delta{\cal A}=0$, with the action ${\cal A}\equiv\int {\cal L}dt$.
So, variation by $\delta\psi$ gives us the first equation of motion ---
the kinematic condition on the free surface
\begin{equation}\label{kinematic_1}
\mbox{Im\,}(Z_t\bar Z_u)=({\delta{\cal K}}/{\delta\psi}).
\end{equation}
Let us divide this equation by $|Z_u|^2$ and use analytical 
properties of the function $Z_t/Z_u$. As a result, we obtain the
time-derivative-resolved equation
\begin{equation}\label{kinematic}
Z_t=iZ_u(1+i\hat H )\left[\frac{(\delta{\cal K}/\delta\psi)}
{|Z_u|^2}\right].
\end{equation}
Further, variation of the action ${\cal A}$ by 
$\delta Z$ gives us the second equation of motion:
\begin{equation}
\left[\frac{\psi_u \bar Z_t-\psi_t\bar Z_u}{2i}\right]=
\left(\frac{\delta{\cal K}}{\delta Z}\right)
+\frac{g}{2i}\,\mbox{Im}\Big(Z\Big)\bar Z_u 
-\frac{(\hat H -i)\Lambda}{2i}.
\label{dynamic_1}
\end{equation}
After multiplying Eq.(\ref{dynamic_1}) by $-2i Z_u$ we have
\begin{equation}
[\psi_t+g\,\mbox{Im\,}Z]|Z_u|^2-\psi_u \bar Z_t Z_u
=(i-\hat H)\tilde\Lambda
-2i\left(\frac{\delta{\cal K}}{\delta Z}\right)Z_u,
\label{dynamic_2}
\end{equation}
where $\tilde\Lambda$ is another real function. Taking the imaginary part of
Eq.(\ref{dynamic_2}) and using Eq.(\ref{kinematic_1}), we find 
$\tilde\Lambda$:
\begin{equation}\label{tilde_Lambda}
\tilde\Lambda =\left[\psi_u \frac{\delta{\cal K}}{\delta\psi}\right]
+2\,\mbox{Re}\left[
\left(\frac{\delta{\cal K}}{\delta Z}\right)Z_u\right].
\end{equation}
After that, the real part of Eq.(\ref{dynamic_2}) gives us the  Bernoulli
equation in a general form:
\begin{eqnarray}
\psi_t&=&-g\,\mbox{Im\,}Z
-\psi_u\hat H\left[\frac{(\delta{\cal K}/\delta\psi)}{|Z_u|^2}\right]
\nonumber\\
\label{Bernoulli}
&+&\frac{\mbox{Im}\left((1-i\hat H)
\left[2({\delta{\cal K}}/{\delta Z})Z_u
+ ({\delta{\cal K}}/{\delta\psi})\psi_u\right]\right)}{|Z_u|^2}.
\end{eqnarray}

Equations (\ref{kinematic}) and (\ref{Bernoulli}) completely determine
evolution of the system, provided the kinetic energy functional 
${\cal K}\{\psi,Z,\bar Z\}$ is explicitly given. It should be emphasized that
in our description a general expression for ${\cal K}$ remains unknown.
However, under the conditions $|z_q|\ll 1$, $|\varphi_q|\ll 1$, the potential
$\varphi(u,v,q,t)$ is efficiently expanded into a series on the powers of the
small parameter $\epsilon$:
\begin{equation}\label{varphi_expansion}
\varphi=\varphi^{(0)} +\varphi^{(1)}+\varphi^{(2)}+\dots,\qquad 
\varphi^{(n)}\sim \epsilon^n,
\end{equation}
where $\varphi^{(n+1)}$ can be calculated from $\varphi^{(n)}$, 
and the zeroth-order term $\varphi^{(0)}=\mbox{Re\,}\phi(w,q,t)$ 
is the real part of an easily represented (in integral form) 
analytical function with the boundary condition
$\mbox{Re\,}\phi|_{v=0}=\psi(u,q,t)$.
Correspondingly, the kinetic energy functional will be written in the form
\begin{equation}\label{H_expansion}
{\cal K}={\cal K}^{(0)}+{\cal K}^{(1)}+\dots, \qquad 
{\cal K}^{(n)}\sim \epsilon^n,
\end{equation}
where ${\cal K}^{(0)}\{\psi\}$ is the kinetic energy of a purely 2D flow,
\begin{eqnarray} 
{\cal K}^{(0)}\{\psi\}&=&\frac{1}{2}\int 
[(\varphi^{(0)}_u)^2+(\varphi^{(0)}_v)^2]\,du\,dv\,dq\nonumber\\
&=&-\frac{1}{2}\int \psi\hat H \psi_u \,du\,dq,
\label{K_0} 
\end{eqnarray}
and the other terms are corrections due to gradients along $q$. 
Now we are going to calculate a first-order correction ${\cal K}^{(1)}$.

\subsection{First-order corrections}

As a result of the conformal change of two variables, the kinetic energy
functional is determined by the expression
\begin{equation}\label{H_full}
{\cal K}=\frac{1}{2}\int\left[\varphi_u^2+\varphi_v^2
+J({\bf Q}\cdot\nabla\varphi)^2\right]du\, dv \,dq,
\end{equation}
where the Cauchy-Riemann conditions $x_u=y_v$, $x_v=-y_u$ have been taken 
into account, and the following notations are used:
$$
J\equiv|z_u|^2, \qquad 
({\bf Q}\cdot\nabla\varphi)\equiv a\varphi_u+b \varphi_v+\varphi_q,
$$
$$
 a=\frac{x_v y_q-x_q y_v}{J}\sim\epsilon^{1/2}, 
\quad b=\frac{y_u x_q-y_q x_u}{J}\sim\epsilon^{1/2}.
$$
Consequently, the Laplace equation in the new coordinates takes the form
\begin{equation}\label{Laplace_uvq}
\varphi_{uu}+\varphi_{vv}
+\nabla\cdot({\bf Q}J({\bf Q}\cdot\nabla\varphi))=0,
\end{equation}
with the boundary conditions
\begin{equation}
\varphi|_{v=0}=\psi(u,q,t),\quad \varphi|_{v=-\infty}=0.
\end{equation}
In the limit $\epsilon\ll 1$ it is possible to write the solution as the
series (\ref{varphi_expansion}), with the zeroth-order term satisfying
the 2D Laplace equation
$\varphi_{uu}^{(0)}+\varphi_{vv}^{(0)}=0,$ 
$\varphi|^{(0)}_{v=0}=\psi(u,q,t)$.
Thus, it can be represented as $\varphi^{(0)}=\mbox{Re\,}\phi(w,q,t)$, where
\begin{equation}\label{phi}
\phi(w,q,t)=\int_{-\infty}^{+\infty}(1-\mbox{sign\,}k)
\psi_k(q,t)e^{ikw}\frac {dk}{2\pi},
\end{equation}
and $\psi_k(q,t)\equiv\int\psi(u,q,t)e^{-iku}du.$
On the free surface
\begin{equation}\label{Psi_def}
\phi(u+i0,q,t)\equiv\Psi(u,q,t)=(1+i\hat H)\psi(u,q,t).
\end{equation}
For all the other terms in Eq.(\ref{varphi_expansion}) we have the relations
\begin{equation}\label{recur}
\varphi_{uu}^{(n+1)}+\varphi_{vv}^{(n+1)}
+\nabla\cdot({\bf Q}J({\bf Q}\cdot\nabla\varphi^{(n)}))=0
\end{equation}
and the boundary conditions $\varphi^{(n+1)}|_{v=0}=0$.
Noting that 
$\int(\varphi_u^{(0)}\varphi_u^{(1)}+\varphi_v^{(0)}\varphi_v^{(1)})\,du\,dv=0$
(it is easily seen without explicit calculation of $\varphi^{(1)}$ 
after integration by parts), we have in the first approximation
\begin{eqnarray}
{\cal K}^{(1)}&=&\frac{1}{2}\int
J(\varphi_q^{(0)}+a\varphi_u^{(0)}+b \varphi_v^{(0)})^2du\, dv \,dq
\nonumber\\
&=&\frac{1}{2}\int
z_u\bar z_u\left[\mbox{Re}\left(\phi_q-\frac{\phi_u z_q}{z_u}\right)
\right]^2du \,dv\, dq.
\label{K1a}
\end{eqnarray}
Since  $z(w)$ and $\phi(w)$ are represented as
$z(u+iv)=e^{-\hat k v}Z(u)$ and $\phi(u+iv)=e^{-\hat k v}\Psi(u)$, 
we can use for $v$-integration the following auxiliary formulas:
\begin{eqnarray}
&& \!\int du\int_{-\infty}^{0} [e^{-\hat k v}A(u)]
\overline{[e^{-\hat k v}B(u)]}\,dv\!
=-\int_{-\infty}^{0} \frac{A_k\overline{B_k}}{2k}\frac{dk}{2\pi}
\nonumber\\
&& 
 \!=-\frac{i}{2}\!\int\! 
\left(\overline{B(u)}\,\hat\partial_u^{-1} \!A(u)\right)du. 
\end{eqnarray} 
Now we apply the above formulas to appropriately decomposed Eq.(\ref{K1a}) 
and, as a result, we obtain an expression of a form
${\cal K}^{(1)}={\cal F}\{\Psi,\overline{\Psi},Z,\overline{Z}\}$,
where the functional ${\cal F}$ is defined as follows 
(compare with \cite{R2005PRE})
\begin{eqnarray}
\!\!{\cal F}\!\!&=&\!\frac{i}{8}
\int(Z_u\Psi_q-Z_q\Psi_u)\hat\partial_u^{-1}
\overline{(Z_u\Psi_q-Z_q\Psi_u)}\,du\,dq
\nonumber\\
&+&\frac{i}{16}\int\!\Bigg\{\!\!\left[
(Z_u\Psi_q-Z_q\Psi_u)^2/{Z_u}\right]\overline{(Z-u)} 
\nonumber\\
&&\qquad - (Z-u)\,\overline{\left[(Z_u\Psi_q-Z_q\Psi_u)^2/{Z_u}\right]}
\!\Bigg\}\,du\,dq.
\label{K_1}
\end{eqnarray}
Here the combination $(Z-u)$ is written instead of $Z$ just for convenience, 
as it is finite at the infinity. Actually the equations of motion 
``do not feel'' this difference. In the derivation we have used the identity
$\int [(Z_u\Psi_q-Z_q\Psi_u)^2/{Z_u}] u du =0$, which holds because the 
integrand is analytical in the lower half-plane.
From here one can express the variational derivatives 
$(\delta{\cal K}^{(1)}/{\delta\psi})$ and $(\delta{\cal K}^{(1)}/{\delta Z})$ 
by the formulas
\begin{equation}\label{derivativesK1}
\frac{\delta{\cal K}^{(1)}}{\delta\psi}=
2\,\mbox{Re\,}\left[(1-i\hat H)\frac{\delta{\cal F}}{\delta\Psi}
\right], \qquad
\frac{\delta{\cal K}^{(1)}}{\delta Z}=
\frac{\delta{\cal F}}{\delta Z}.
\end{equation}
The derivatives
$(\delta{\cal F}/{\delta\Psi})$ and $(\delta{\cal F}/{\delta Z})$
are calculated in a standard manner:
\begin{eqnarray}
\frac{\delta{\cal F}}{\delta\Psi}&=&\frac{i}{8}
Z_q\Big[\overline{(Z_u\Psi_q-Z_q\Psi_u)} \nonumber\\
&&\qquad\qquad+\hat\partial_u[(\Psi_q-Z_q\Psi_u/{Z_u})
\overline{(Z-u)}]\Big]
\nonumber\\
&&-\frac{i}{8}\,Z_u\,\hat\partial_q\Big[
\hat\partial_u^{-1}\overline{(Z_u\Psi_q-Z_q\Psi_u)}\nonumber\\
&&\qquad\qquad + (\Psi_q-Z_q\Psi_u/{Z_u})\overline{(Z-u)}\Big],
\end{eqnarray}

\begin{eqnarray}
\frac{\delta{\cal F}}{\delta Z}&=&-\frac{i}{8}
\Psi_q\Big[\overline{(Z_u\Psi_q-Z_q\Psi_u)}\nonumber\\
&&\qquad\qquad
+ \hat\partial_u[(\Psi_q-Z_q\Psi_u/{Z_u})\overline{(Z-u)}]\Big]
\nonumber\\
&&+\frac{i}{8}\,
\Psi_u\,\hat\partial_q\Big[
\hat\partial_u^{-1}\overline{(Z_u\Psi_q-Z_q\Psi_u)}\nonumber\\
&&\qquad\qquad+ (\Psi_q-Z_q\Psi_u/{Z_u})\overline{(Z-u)}\Big]\nonumber\\
&&+\frac{i}{16}\,\Big[\hat\partial_u[(\Psi_q-Z_q\Psi_u/Z_u)^2
\overline{(Z-u)}]\nonumber\\
&&\qquad\qquad-\overline{(\Psi_q-Z_q\Psi_u/{Z_u})^2{Z_u}}
\Big].
\end{eqnarray}
Now one can  substitute $(\delta{\cal K}/\delta\psi)\approx -\hat H\psi_u
+(\delta{\cal K}^{(1)}/\delta\psi)$ and 
$(\delta{\cal K}/\delta Z)\approx(\delta{\cal K}^{(1)}/\delta Z)$
into the equations of motion (\ref{kinematic}) and (\ref{Bernoulli}).
Thus, the required weakly 3D equations for deep water waves are completely 
derived:
\begin{equation}\label{Z_Y}
Z=u+(i-\hat H)Y(u,q,t), \quad Z_u=1+(i-\hat H)Y_u.
\end{equation}
\begin{equation}\label{kinematic_deep}
Z_t=-iZ_u(1+i\hat H)\left[
{[\hat H\psi_u-(\delta{\cal K}^{(1)}/\delta\psi)]}/
{|Z_u|^2}\right],
\end{equation}
\begin{eqnarray}
\psi_t\!\!&=&\!-g\,Y+
\,\psi_u\hat H\left[{[\hat H\psi_u-(\delta{\cal K}^{(1)}/
\delta\psi)]}/{|Z_u|^2}\right]\nonumber\\
&+&\!{\hat H\left[\psi_u [\hat H\psi_u-(\delta{\cal K}^{(1)}/
{\delta\psi})]\right]}/{|Z_u|^2}\nonumber\\
\label{Bernoulli_deep}
&-&\!{2\,\mbox{Re}\left((\hat H+i)
[Z_u({\delta{\cal K}^{(1)}}/{\delta Z})]\right)}/{|Z_u|^2}.
\end{eqnarray}

\subsection{Modification of the Hamiltonian}

It can be easily obtained that the linear dispersion relation for the system
(\ref{K_1})-(\ref{Bernoulli_deep}) is
\begin{equation}\label{omega_approx}
\omega^2(k,m)=g|k|\left(1+\frac{1}{2}\frac{m^2}{k^2}\right),
\end{equation}
where $m$ is the wave-number in the $q$-direction (the wave-number $k$ in the 
$u$-direction was introduced earlier). Obviously, here we have the first two
terms from the expansion of the exact 3D linear dispersion relation
\begin{equation}\label{omega_exact}
\omega^2(k,m)=g\sqrt{k^2+m^2}
\end{equation}
on the powers of $m^2/k^2\ll 1$. Thus, the system 
(\ref{K_1})-(\ref{Bernoulli_deep}) has a non-physical singularity 
in the dispersion relation near the 
$m$-axis in the Fourier plane $(k,m)$, where $k/m\to 0$. 
Therefore, for convenience of the numerical modeling,
some regularization could be useful in the approximate Hamiltonian 
${\cal K}^{(0)}+{\cal K}^{(1)}+{\cal P}$, 
where ${\cal P}$ is the potential energy,
\begin{equation}\label{PotentialEnergy}
{\cal P}=\frac{g}{2}\int\left[\frac{Z-\bar Z}{2i}\right]^2
\left[\frac{Z_u+\bar Z_u}{2}\right]du\,dq.
\end{equation}  
Also, the correct linear dispersion 
relation is strongly desirable, since it determines which waves interact
by nonlinear resonances.
Of course, a possible regularization is not unique as we keep 
only zeroth- and first-order terms on $\epsilon=(m/k)^2$ 
in the Hamiltonian.
Below we suggest a modification which adds terms of the order 
${\cal O}(\epsilon^2)$ to the approximate Hamiltonian 
${\cal P}+{\cal K}^{(0)}+{\cal K}^{(1)}$ 
(a modified Hamiltonian $\tilde{\cal H}={\cal P}+\tilde{\cal K}$ 
will remain valid up to $\epsilon$). 
First of all, instead of the the functional ${\cal K}^{(0)}$ 
we use another functional, 
${\cal K}^{(0)}\to(1/2) \int \psi\hat G_0\psi \,du\,dq$, 
where the linear operator $\hat G_0$ is diagonal in the Fourier representation:
\begin{equation}\label{G0_modified}
G_0(k,m)=\sqrt{k^2+m^2}-\frac{1}{2}\frac{|k| m^2}{k^2+m^2}.
\end{equation}
Besides that, we change the operator $i\hat\partial_u^{-1}=1/k$ 
in the first line of Eq.~(\ref{K_1}) by the  operator 
$k/(k^2+m^2)=i\hat\partial_u\Delta_2^{-1}$, which is less singular.
As the result, we have the modified approximate kinetic-energy functional 
in the form
\begin{equation}
\tilde{\cal K}=\frac{1}{2}\int \psi\hat G_0\psi \,du\,dq +\tilde{\cal F},
\end{equation}
\begin{eqnarray}
\tilde{\cal F}&=&\frac{i}{8}
\int(Z_u\Psi_q-Z_q\Psi_u)\hat\partial_u\Delta_2^{-1}
\overline{(Z_u\Psi_q-Z_q\Psi_u)}\,du\,dq
\nonumber\\
&+&\frac{i}{16}\int\Bigg\{\left[
(Z_u\Psi_q-Z_q\Psi_u)^2/{Z_u}\right]\overline{(Z-u)} 
\nonumber\\
&&\qquad - (Z-u)\,\overline{\left[(Z_u\Psi_q-Z_q\Psi_u)^2/{Z_u}\right]}
\Bigg\}\,du\,dq.
\label{H_modified}
\end{eqnarray}
The linear dispersion relation resulting from $\tilde{\cal K}$ is correct 
in the whole Fourier plane. Besides that,
the zeroth- and the first-order terms on $\epsilon$ in $\tilde{\cal K}$
are the same as in ${\cal K}^{(0)}+{\cal K}^{(1)}$.
The system of equations, corresponding to the modified Hamiltonian 
$\tilde{\cal H}$, consists of Eq.~(\ref{Z_Y}) and the following equations:
\begin{equation}\label{kinematic_deep_modified}
Z_t=iZ_u(1+i\hat H)\left[
{(\delta\tilde{\cal K}/\delta\psi)}/{|Z_u|^2}\right],
\end{equation}
\begin{eqnarray}
\psi_t&=&{\mbox{Im}\left((1-i\hat H)
[2({\delta\tilde{\cal F}}/{\delta Z})Z_u+
(\delta\tilde{\cal K}/\delta\psi)\psi_u]\right)}/{|Z_u|^2}
\nonumber\\
\label{Bernoulli_deep_modified}
&-&g\,Y
-\psi_u\hat H\left[(\delta\tilde{\cal K}/\delta\psi)/|Z_u|^2\right],
\end{eqnarray}
\begin{equation}\label{tilde_H_psi}
\frac{\delta\tilde{\cal K}}{\delta\psi}=\hat G_0\psi+
2\,\mbox{Re\,}\left[(1-i\hat H)\frac{\delta\tilde{\cal F}}{\delta\Psi}
\right],
\end{equation}

\begin{eqnarray}
\frac{\delta\tilde{\cal F}}{\delta\Psi}&=&\frac{i}{8}
Z_q\hat\partial_u\Big[\hat\partial_u\Delta_2^{-1}
\overline{(Z_u\Psi_q-Z_q\Psi_u)} \nonumber\\
&&\qquad\qquad+(\Psi_q-Z_q\Psi_u/{Z_u})
\overline{(Z-u)}\Big]
\nonumber\\
&&-\frac{i}{8}\,Z_u\,\hat\partial_q\Big[
\hat\partial_u\Delta_2^{-1}\overline{(Z_u\Psi_q-Z_q\Psi_u)}\nonumber\\
&&\qquad\qquad + (\Psi_q-Z_q\Psi_u/{Z_u})\overline{(Z-u)}\Big],
\label{tilde_F_Psi}
\end{eqnarray}
\begin{eqnarray}
\frac{\delta\tilde{\cal F}}{\delta Z}&=&-\frac{i}{8}
\Psi_q\hat\partial_u\Big[\hat\partial_u\Delta_2^{-1}\overline{(Z_u\Psi_q-Z_q\Psi_u)}\nonumber\\
&&\qquad\qquad
+ (\Psi_q-Z_q\Psi_u/{Z_u})\overline{(Z-u)}\Big]
\nonumber\\
&&+\frac{i}{8}\,
\Psi_u\,\hat\partial_q\Big[
\hat\partial_u\Delta_2^{-1}\overline{(Z_u\Psi_q-Z_q\Psi_u)}\nonumber\\
&&\qquad\qquad+ (\Psi_q-Z_q\Psi_u/{Z_u})\overline{(Z-u)}\Big]\nonumber\\
&&+\frac{i}{16}\,\Big[\hat\partial_u[(\Psi_q-Z_q\Psi_u/Z_u)^2
\overline{(Z-u)}]\nonumber\\
&&\qquad\qquad\qquad-\overline{(\Psi_q-Z_q\Psi_u/{Z_u})^2{Z_u}}
\Big].
\label{tilde_F_Z}
\end{eqnarray}

These equations were integrated numerically as described in the 
following section. It is interesting to note that we also considered another 
choice for the regularization, with $G_0=(k^2+m^2/2)(k^2+m^2)^{-1/2}$ 
instead of $|k|$ in ${\cal K}^{(0)}$, and $(k^2+m^2)^{-1/2}$ instead 
of $1/|k|$ in the first term in ${\cal K}^{(1)}$.
The numerical results were found very close in both cases, since the wave
spectra were concentrated in the region $m^2/k^2\ll 1$.

\section{Numerical method and results}

In our numerical simulations, we used the following procedure.
A rectangular domain $L_u\times L_q$ with the periodic boundary conditions 
in the $(u,q)$-plane was reduced to the standard dimensionless size 
$2\pi\times 2\pi$ [the aspect ratio $\varepsilon=(L_u/L_q)^2$ was taken 
into account]. This standard square was discretized by $N\times L$ points 
$(u_n,q_l)=2\pi(n/N,l/L)$, with integer $n$ and $l$. 
The time variable was rescaled to give $g=1$.
As the primary dynamical variables, the (complex) Fourier components 
$Y_{km}$ and $P_{km}=(k^2+\varepsilon m^2)^{1/4}\psi_{km}$ 
were used, where $k$ and $m$ were integer numbers in the limits $0\le k<K$, 
$-M\le m\le M$, with $K\approx 3N/8$, $M\approx 3L/8$.
For negative $k$, the properties $Y_{-k,-m}=\overline{Y_{km}}$ and 
$P_{-k,-m}=\overline{P_{km}}$ were implied.
It should be noted that after rescaling  $u$, $q$, and  $t$,
the linear dispersion relation has been 
$\omega_{km}=(k^2+\varepsilon m^2)^{1/4}$, and that the combinations 
$$
a_{km}=\frac{Y_{km}+iP_{km}}{\sqrt{2\omega_{km}}}
$$
in the linear limit coincide with the normal complex variables 
\cite{DLZ95,Z1999}.

Using the FFTW library \cite{fftw3}, the quantities
$Z(u_n,q_l)$, $\Psi(u_n,q_l)$, their corresponding $u$- and 
$q$-derivatives, and $\hat G_0\psi(u_n,q_l)$ are represented 
by 2-dimensional complex arrays.
The variational derivatives (\ref{tilde_F_Psi}), (\ref{tilde_F_Z}),
(\ref{tilde_H_psi}) and the right-hand-sides of 
Eqs.~(\ref{kinematic_deep_modified})-(\ref{Bernoulli_deep_modified})
are then computed and transformed into Fourier space in order to
evaluate the multi-dimensional function
which determines the time derivatives of $Y_{km}$ and $P_{km}$.
This function is used in the standard fourth-order Runge-Kutta 
procedure  (RK4) for the time integration of the system. 
After each RK4 step, a large-wave-number filtering of the arrays 
$Y_{km}$ and $P_{km}$ is carried out, so that only the Fourier components with 
$0\le k<K_{eff}$, $-M_{eff}\le m\le M_{eff}$ are kept, 
where $K_{eff}\approx N/4$ and $M_{eff}\approx L/4$. 
To estimate the accuracy of the computations, conservation of the total energy 
$\tilde{\cal H}$ and the mean surface elevation are monitored. 

Typically, at $t=0$ we put $N=2^{12}$, $L=2^{6}$, and the initial time step
$\tau=0.01$. During the computation, as wave crests become more sharp
and the spectra get broadened, 
we several times adaptively double $N$ (together with $K$, $K_{eff}$) 
and $L$ (together with $M$, $M_{eff}$), with
the time step half-decreasing when $N$ is doubled. 
At the end, when a giant wave is formed, we have $N=2^{13-14}$, $L=2^{7-8}$.
As a result of such adaptive scheme, the conservation of the total energy
is kept up to 5-6 decimal digits during most part of the evolution. 
Only at a very late stage, when $N$ and $L$ are not allowed to double
anymore, the conservation is just up to 3-4 digits, 
and the filtering of the higher harmonics becomes more influential.

The efficiency of the above described numerical method crucially depends on 
the speed of the FFT routine, since most of the computational
time (approximately 80\%) is spent in the Fourier transforms.
To compute the evolution through the unit (dimensionless) time period, 
on a modern PC (Intel Pentium IV 3.2 GHz) it takes 
three-four minutes with $N=4096$, $L=64$, $\tau=0.01$,
and more than one hour with $N=8192$, $L=256$, $\tau=0.005$.
The total time needed for a single numerical rogue-wave experiment
is about 3-5 days. 
\begin{figure}
\begin{center} 
  \epsfig{file=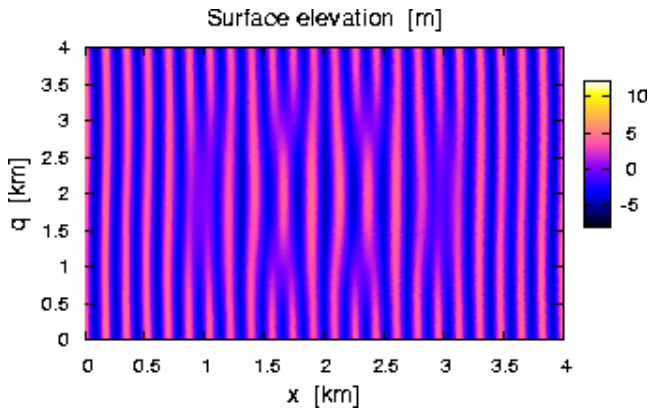,width=85mm}
\end{center}
\caption{(Color online) [A]. Map of the free surface at $t=0$.} 
\label{A-t0-XY}
\end{figure}
\begin{figure}
\begin{center} 
   \epsfig{file=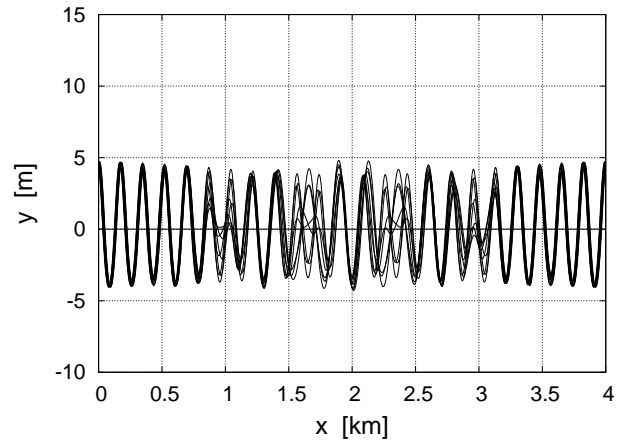,width=85mm} 
\end{center}
\caption{[A]. Initial wave profiles
for eight equidistant values $q=L_q(0,1/8, \dots, 7/8)$. } 
\label{A-t0-profiles}
\end{figure}
\begin{figure}
\begin{center} 
    \epsfig{file=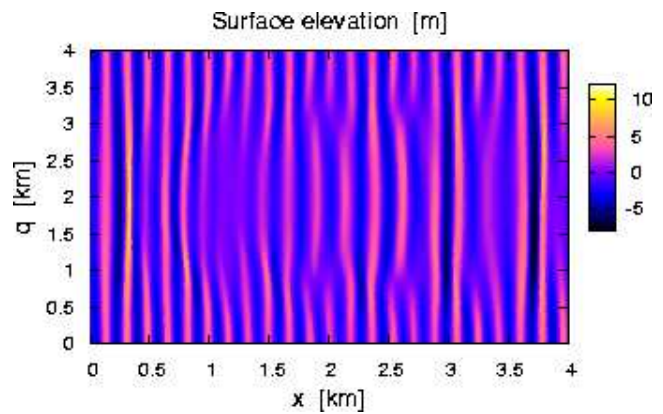,width=85mm}      
\end{center}
\caption{(Color online) [A]. Map of the free surface at $t=59.2$ 
(the physical time is 7 min 57 sec).} 
\label{A-t1-XY}
\end{figure}
\begin{figure}
\begin{center} 
   \epsfig{file=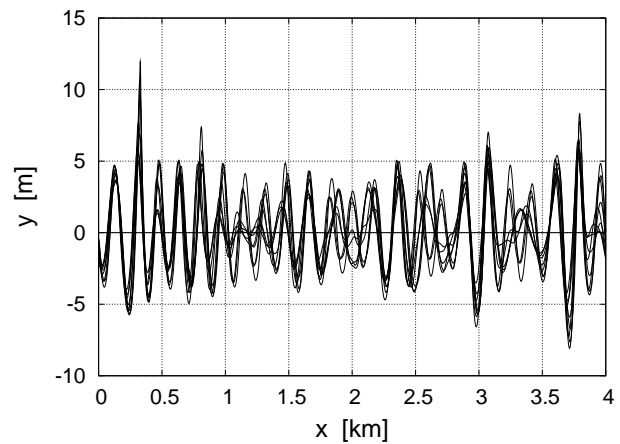,width=85mm}     
\end{center}
\caption{[A]. Wave profiles at $t=59.2$. } 
\label{A-t1-profiles}
\end{figure}

Two numerical experiments are reported below for horizontal physical box 
sizes of $4\times 4$ km (the final resolution was $16384\times 256$ in both). 
In the first of them, the main mechanism for a big wave formation 
is the linear dispersion, when a group of
longer waves overtakes a group of shorter waves, and their amplitudes are 
added (see \cite{Kharif-Pelinovsky} for a discussion). 
However, due to nonlinear effects, in maximum the 
crest is higher than simply the sum of two amplitudes. 
In the second experiment, a giant wave is formed by an essentially nonlinear
mechanism (due to Benjamin-Feir instability) from a slightly modulated 
periodic wave, close to a stationary Stokes wave. 

\subsection{``Linear'' big wave}

In the first numerical experiment (referred to as [A]), the initial state was 
a composition of two spatially separated wave groups, as shown in 
Fig.\ref{A-t0-XY}. 
 Two typical dimensionless wave numbers are present, 
$k_1=17$, corresponding to the wave length $\lambda_1\approx 235$ m 
(at the central part of Fig.~\ref{A-t0-XY}) and $k_2=23$ 
($\lambda_2\approx 174$ m). 
At $t=0$ the highest crests were about 5 m in both groups 
(see Fig.~\ref{A-t0-profiles}). 
Due to  difference in the group velocities ($c_{gr}\sim|k|^{-1/2}$), 
the longer waves move faster and after some time overtake the shorter waves.
As a result of almost linear superposition, big waves with high crests 
and deep troughs are formed, separated by the spatial period $2\pi/|k_2-k_1|$,
with the maximum amplitude about 12 m, which is approximately equal to the 
sum of the individual amplitudes (the nonlinearity slightly increases the 
maximum height). The corresponding numerical results are presented 
in Figs.\ref{A-t1-XY} and \ref{A-t1-profiles}.

\subsection{Nonlinear big wave}

\begin{figure}
\begin{center}
   \epsfig{file=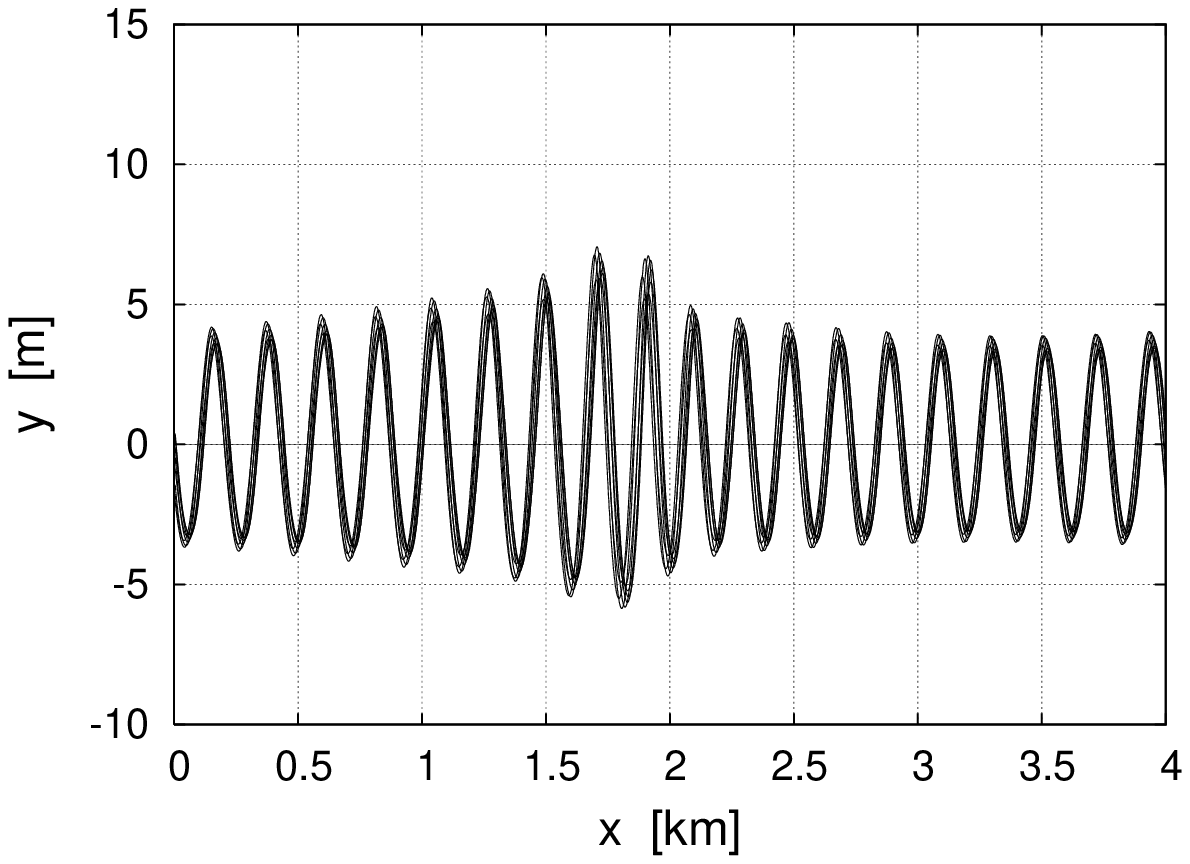,width=42mm}
   \epsfig{file=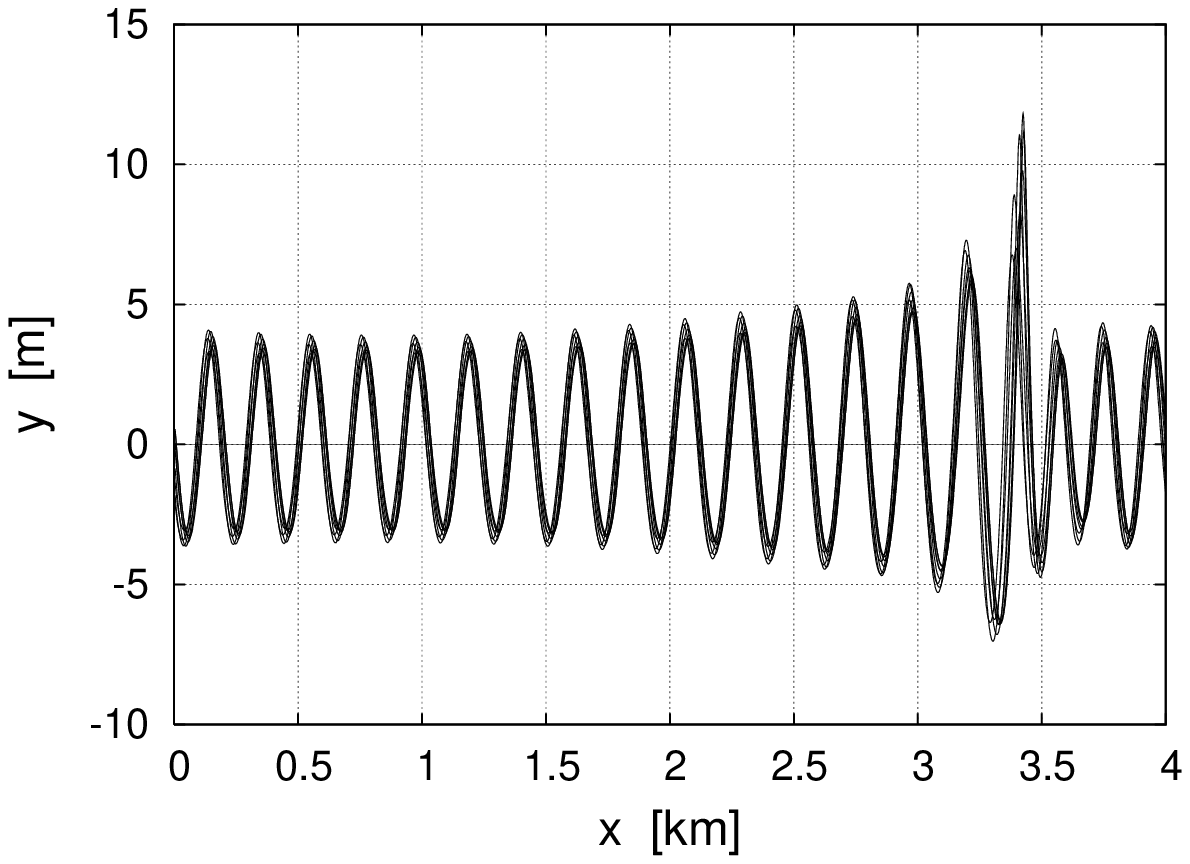,width=42mm} 
\end{center}
\caption{[B]. Wave profiles at $t=150$ (20 min 9 sec), 
and at $t=170$ (22 min 50 sec).} 
\label{B-t1-t2-profiles}
\end{figure}
\begin{figure}
\begin{center} 
     \epsfig{file=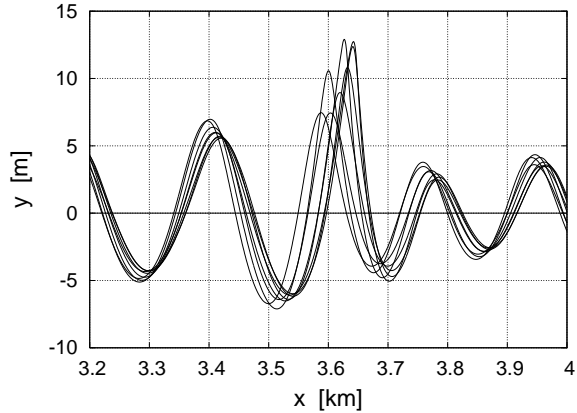,width=80mm}   
\end{center}
\caption{[B]. Rogue wave at $t=172.8$ (23 min 13 sec).} 
\label{B-t3-profiles}
\end{figure}
\begin{figure}
\begin{center} 
   \epsfig{file=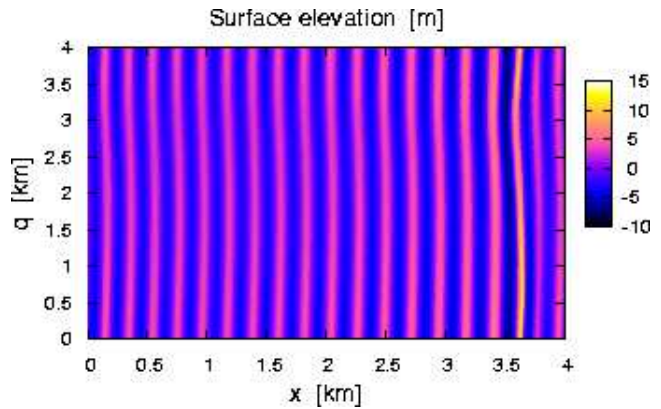,width=85mm}      
\end{center}
\caption{(Color online) [B]. Map of the free surface at $t=172.8$.}
\label{B-t3-XY}
\end{figure}

Our second  numerical experiment (referred to as [B]) is a weakly 3D analog of
the 2D numerical experiments performed by Zakharov and co-workers 
\cite{ZDV2002,DZ2005Pisma}, when a slightly modulated periodic Stokes wave 
evolves to produce a giant wave. However, with two horizontal dimensions 
we could not achieve the same very high resolution as Zakharov with co-workers 
did for the case of a single horizontal dimension ($16384\times 256$ in our 
experiment [B] versus $2\cdot10^6\times 1$ in Ref.~\cite{DZ2005Pisma}). 
Our computations were terminated well before the 
moment when a giant wave reached its maximum height and began break, 
since the number of the employed Fourier modes became inadequate. 
The accuracy could be better with larger $N$ and $L$, and with a smaller 
$\tau$, but it required much more memory and computational time.
In general, to resolve in conformal variables a sharp wave crest with a 
minimal radius of curvature  $\rho$ and with the asymptotic angle $2\pi/3$ 
(as in the limiting Stokes wave), the required number $K_{eff}$ 
should adaptively vary as $\rho^{-3/2}\lambda^{1/2}$. The power $-3/2$ 
results in strong difficulties when $\rho$ is small.
Another important point is that Zakharov and co-workers were able to 
re-formulate the purely 2D equations in terms of the ``optimal'' 
complex variables $R=1/Z_u$ and $U=i\Psi_u/Z_u$, thus obtaining very elegant
and compact cubic evolution equations (the Dyachenko equations \cite{D2001}). 
In our 3D case, a similar simplification seems to be impossible, 
and we dealt directly with the original conformal variables $Z$ and $\Psi$.
Nevertheless, our results are sufficiently accurate 
to reproduce the fact of a giant wave formation.

As the initial state for the experiment [B], we took a weakly modulated 
periodic wave with the main wave number $k=19$ ($\lambda\approx 210$ m) 
and with a few first harmonics on $2k$, $3k$, etc., similar to a Stokes wave
(not shown). After some period of evolution, the Benjamin-Feir instability 
developed and resulted in formation of a big wave, with the amplitude  
13 m at $t=172.8$ versus the initial maximum amplitude 5 m
(see Figs.~\ref{B-t1-t2-profiles}-\ref{B-t3-XY} and compare with 
Ref.~\cite{DZ2005Pisma}). The peak-to-trough height $h_*$ of this computed 
rogue wave was approximately 20 m at $t=172.8$ (the steepness parameter 
$h_*/\lambda\approx 0.1$), and it was still growing at that moment
(so, at $t=173.0$ we observed the amplitude 14 m, but the accuracy was already
not sufficient).
It is interesting that this numerical solution has a well defined envelope 
until a very final stage of evolution. Thus, our equations may serve to test 
the simplified wave-packet models like the extended NLS equations
\cite{Dysthe1979,TKDV2000,OOS2000}.

\section{Summary and discussion}

We have developed an efficient numerical method for modeling the rogue wave 
phenomenon. The underlying theory for the method is the weakly 3D
formulation of the free surface dynamics reported earlier \cite{R2005PRE} with
slight modifications. In particular, the Hamiltonian has been regularized in a 
way to give the exact linear 2D dispersion relation in the entire Fourier space, 
and a filter removing shortest waves has been added in the numerical 
implementation. With these techniques, weakly three-dimensional effects could 
be included in simulations of rogue wave formation as illustrated in the two 
examples given. In particular, the genuinely non-linear 2D instability reported 
by Zakharov et al. \cite{ZDV2002,DZ2005Pisma} could be verified in the weakly 
3D regime despite the relatively low resolution (compared to 2D) of 
$16384\times 256$ points used here.

The results indicate that the assumption of weak variation in the
third direction holds even in the late stage of rogue wave formation,
which demonstrates the consistency of the expansion in
$\epsilon=(l_x/l_q)^2$ and thereby the applicability of the
present theory.

Planned further steps in the continuation of this work are a more
efficient computational implementation through parallelization of the
code and the inclusion of additional effects like a bottom profile,
which is already covered by the formalism reported earlier
\cite{R2005PRE}.


\end{document}